\documentclass[preprint,11pt,aps,draft,nofootinbib,showpacs,showkeys]{revtex4}
\usepackage{amssymb}

\newtheorem{tm}{Theorem}
\newtheorem{de}{Definition}
\newtheorem{pr}{Proposition}

\begin{document}

\title{Quantum mechanical virial theorem in systems with translational and rotational symmetry}

\author{Domagoj Kui\'{c}}

\email{dkuic@pmfst.hr}

\affiliation{University of Split, Faculty of Science,\\ N. Tesle 12, 21000 Split, Croatia}

\date{December 6, 2012}

\begin{abstract}
Generalized virial theorem for quantum mechanical nonrelativistic and  relativistic systems with translational and rotational symmetry is derived in the form of the commutator between the generator of dilations $G$ and the Hamiltonian $H$. If the conditions of translational and rotational symmetry together with the additional conditions of the theorem are satisfied, the matrix elements of the commutator $[G, H]$ are equal to zero on the subspace of the Hilbert space. Normalized simultaneous eigenvectors of the particular set of commuting operators which contains $H$, $\mathbf J^2, \ J_{z}$ and additional operators form an orthonormal basis in this subspace. It is expected that the theorem is relevant for a large number of quantum mechanical $N$-particle systems with translational and rotational symmetry. 
\end{abstract}


\maketitle

\section{Introduction}

As is the case with many interesting quantum mechanical systems, bound states seem beyond standard perturbation theory and are difficult to grasp, even more so because very few systems are known to be solvable exactly. Considering the limitations, virial theorem remains a very useful tool in studying bound states and quantum mechanical systems in general. Historically, derivation of the quantum mechanical virial theorem has been led by analogy with the classical counterpart. Quantum mechanical virial theorem in the nonrelativistic form is the relation between expectation values of the kinetic energy and the directional derivative of the potential energy. For a system described by a relativistic Hamiltonian, quantum mechanical virial theorem is the relation between expectation values of directional derivatives of the kinetic and potential energy. An account of virial theorems in quantum mechanics, including relativistic and nonrelativistic Hamiltonians and wave equations (Schr\"odinger, Salpeter, Dirac, Klein-Gordon) is found in \cite {1}.          

Relations between virial theorem in classical mechanics and virial theorem in quantum mechanics can be understood on the basis of Ehrenfest's theorem \cite{9} and the correspondence principle. Derivations of the classical and quantum mechanical virial theorems can be found in standard classical and quantum mechanics textbooks, e.g. \cite {2,3}. Quantum mechanical virial theorem is derived from the time derivative of the expectation value of operator $G$, i.e. from the expectation value of the commutator between $G$ and $H$. It is a simple fact in quantum mechanics that the expectation value of the commutator $[G, H]$ vanishes for eigenvectors of the Hamiltonian $H$ and the quantum mechanical virial theorem follows. This is explained briefly in Sect. \ref {sec:Vtfcaqs}. It is known that this argument is only formal since the generator of dilations $G$ is an unbounded operator and an eigenvector of $H$ need not be in a domain of operator $G$ \cite[Vol. 4, 231]{4}. Review of the various assumptions under which abstract versions of the quantum mechanical virial theorem have been proved is found in \cite {5}.  Other ways of deriving the quantum mechanical virial theorem include an approach using dilations. Detailed explanations of this approach and other approaches can be found in reference \cite{1}. Some of the results from reference \cite{1} are used here in Sect. \ref {sec:dilations}. There is also an approach using variational calculation \cite {6}. For relativistic particles described by the Dirac or the Klein-Gordon equation and bound in scalar and/or vector type potentials an example of variational approach is given in \cite {7}. 

Section \ref {sec:theorem} represents a possible step further under assumption of certain symmetry properties of the Hamiltonian. In the proof of the theorem in Sect. \ref {sec:theorem} no reference is made if the particular system on which the theorem could be applied is described by a nonrelativistic  or relativistic Hamiltonian. It is required, among other conditions of the theorem in Sect. \ref {sec:theorem}, that the Hamiltonian is invariant to translations and rotations, i.e. that the system described by the Hamiltonian has translational and rotational symmetry. If in addition to translations and rotations, the Hamiltonian is invariant to other symmetry operations, there is an additional condition. This additional condition of the theorem in Sect. \ref {sec:theorem} is important if, in addition to $H,\ \mathbf {J}^{2}$ and $J_{z}$, there is a set of self-adjoint operators $\{\Omega ^\alpha \}$ which are members of the complete set of commuting operators $\{H$, $\mathbf {J}^{2}$, $J_{z}$, $\{\Omega ^\alpha \}\}$. It is then required that all operators $\Omega ^\alpha$ in the set $\{\Omega ^\alpha \}$ commute with the generator of dilations $G$. If this requirement is satisfied, along with translational and rotational symmetry and other conditions of the theorem in Sect. \ref {sec:theorem}, the matrix elements of the commutator $[G, H]$  are equal to zero on the subspace $\mathcal {D}(H, \mathbf {J}^{2}, J_{z}, \{\Omega ^\alpha \})$ of the Hilbert space. The subspace $\mathcal {D}(H, \mathbf {J}^{2}, J_{z}, \{\Omega ^\alpha \})$ is generated by the normalized simultaneous eigenvectors of the set of commuting operators $\{H$, $\mathbf {J}^{2}$, $J_{z}$, $\{\Omega ^\alpha \}\}$ which form an orthonormal basis in this subspace. 

The conditions of the theorem in Sect. \ref {sec:theorem} are not satisfied for example, if a member of the set $\{\Omega ^\alpha \}$ exists which does not commute with the operator $G$. Therefore, possible applications of the theorem in Sect. \ref {sec:theorem} should be considered with this limitation in mind. Another limitation is that an eigenvector of the set of commuting operators $\{H$, $\mathbf {J}^{2}$, $J_{z}$, $\{\Omega ^\alpha \}\}$ need not be in a domain of operator $G$. These and other limitations are discussed in more detail in Sect. \ref {sec:theorem} and at the end of Sect. \ref{sec:dilations}. The proof of the theorem in Sect. \ref {sec:theorem} includes the use of analytic functions and distributions. Part of the proof extends to the Appendix \ref{app:1}. It is expected that the theorem in Sect. \ref {sec:theorem} relates to a large number of nonrelativistic and relativistic quantum mechanical $N$-particle systems with translational and rotational symmetry.

\section{Virial theorem in quantum mechanics}\label {sec:Vtfcaqs}

In the nonrelativistic quantum mechanics and in relativistic quantum mechanics, where the Hamiltonian for the system described by the particular wave equation can be given, virial theorem follows for eigenvectors of the Hamiltonian as a consequence of hermiticity of the Hamiltonian, $H^{\dagger } = H$. Quantum mechanical virial theorem \cite{1} is in this way equivalent to a statement that the expectation values of the commutator $[G, H]$ are equal to zero
\begin{equation}
\label {eq:VT_sec2} \langle \psi  \vert [G, H] \vert \psi  \rangle = 0 ,
\end{equation}
taken with respect to the normalized eigenvectors of the Hamiltonian $H$, 
\begin{eqnarray}
H \vert \psi \rangle = E\vert \psi  \rangle ,&\qquad  \qquad & \langle \psi  \vert \psi  \rangle = 1 .  
\end{eqnarray}
This argument is formal since the generator of dilations $G$ is an unbounded operator and $\vert \psi \rangle $ need not be in a domain of operator $G$, \cite [Vol. 4, 231]{4}. For $N$-particle system operator $G$ is given by
\begin{equation}
\label {eq:opGh} G = \frac{1}{2}\sum _{i=1}^{N}\left (\mathbf {r}_{i}\cdot \mathbf {p}_{i} + \mathbf {p}_{i}\cdot \mathbf {r}_{i} \right ) .
\end{equation}

\section{Systems with translational and rotational symmetry}\label {sec:theorem}

\subsection{Symmetry conditions}

In Sect. \ref {sec:Vtfcaqs} quantum mechanical virial theorem was given in the form that is common in the literature for the nonrelativistic and relativistic cases. This form of quantum mechanical virial theorem directly depends on the state vectors; the proof is formal and is given only for eigenvectors of the Hamiltonian. It will be shown here in Sect. \ref{sec:theorem}, that under the conditions of translational and rotational symmetry of the Hamiltonian and the additional conditions, the virial theorem derived in the form of the commutator between the generator of dilations $G$ and the Hamiltonian $H$ is true on the subspace of the Hilbert space, because the matrix elements of this commutator are equal to zero on this subspace of the Hilbert space. A formal proof of this generalized quantum mechanical virial theorem is given here without specific reference to the detailed form of the Hamiltonian. It should work for the nonrelativistic and relativistic Hamiltonians, if along with other conditions of the theorem, the fundamental conditions of translational and rotational symmetry are satisfied, i.e. the following commutators are equal to zero
\begin{eqnarray}
\label {eq:HcommJp} [H, \mathbf {P}]=0 , &\qquad \qquad & [H, \mathbf {J}]=0 .
\end{eqnarray}
{\it Other conditions additional} to (\ref {eq:HcommJp}) exist, and will be discussed. The generator of translations, operator of the total momentum $\mathbf {P}=\sum_{i=1}^{N}\mathbf {p}_{i}$ and the generator of rotations, operator of the total angular momentum $\mathbf {J}=\sum_{i=1}^{N}\mathbf {J}_{i}$, do not commute. Possible set of commuting operators can contain operators $H,\ \mathbf {J}^{2}$ and $\ J_{z}$, or it can contain operators $H$ and $\mathbf {P}$, but can not contain operators $H,\ \mathbf {J}^{2}, \ J_{z}$ and $\mathbf P$. 

To begin, the properties of operators $G ,\ H$ and $\mathbf {J}$ are discussed first. As given by Eq. (\ref {eq:opGh}), $G$ is a scalar operator and therefore is invariant to rotations. This means that $G$ commutes with the generator of rotations, operator of the total angular momentum $\mathbf {J}=\sum_{i=1}^{N}\mathbf {J}_{i}$. The commutators of the operator $G$ with the operators $\mathbf {J}$ and $\mathbf {P}$ are equal to
\begin{eqnarray}
\label {eq:GcommJpp} [G, \mathbf {J}] = 0 , & \qquad \qquad & [G, \mathbf {P}] = i\hbar \mathbf {P} .  
\end{eqnarray}   

The discussion in this section is limited to quantum mechanical $N$-particle systems that have the following symmetry properties:

\begin{de}\label{definition1}
The Hamiltonian $H$ of the system considered here is invariant to translations, rotations and can be invariant to additional symmetry operations. The commutators of the Hamiltonian $H$ with the generators of translations and rotations, operators $\mathbf P$ and $\mathbf J$, are equal to zero. If, in addition to $\mathbf P$ and $\mathbf J$, other self-adjoint operators which commute with $H$ exist, then the additional condition is required from these operators. Those of them which commute with $\mathbf {J}^2, \ J_z$ and among themselves, and hence are members of the set of commuting operators also containing  $H,\ \mathbf {J}^{2}$ and $\ J_{z}$, should commute with the operator $G$.
\end{de}

Suppose that in addition to $\mathbf P$ and $\mathbf J$, other self-adjoint operators which commute with $H$ exist. Suppose that they satisfy the aforementioned additional condition. The set $\{\Omega ^\alpha \}$ is formed by all self-adjoint operators $\Omega ^\alpha $ satisfying the following relations required by Definition \ref{definition1}:
\begin{eqnarray}
& [H, \Omega ^\alpha ] = 0 , & \qquad \quad \forall \Omega ^\alpha \in \{\Omega ^\alpha \} , \nonumber\\
& [\mathbf J^2, \Omega ^\alpha ] = 0 , & \qquad \quad \forall \Omega ^\alpha \in \{\Omega ^\alpha \} , \nonumber\\
& [J_z, \Omega ^\alpha ] = 0 , & \qquad \quad \forall \Omega ^\alpha \in \{\Omega ^\alpha \} , \nonumber\\
& [\Omega ^\alpha , \Omega ^{\alpha ^\prime} ] = 0 , & \qquad \quad \forall \Omega ^\alpha , \forall \Omega ^{\alpha ^\prime} \in \{\Omega ^\alpha \} , 
\end{eqnarray}
and since the additional condition is fulfilled,
\begin{eqnarray}
\label {eq:GcommOmega} & [G, \Omega ^\alpha ] = 0 , & \qquad \quad \forall \Omega ^\alpha \in \{\Omega ^\alpha \} . 
\end{eqnarray}
For example, if the Hamiltonian $H$, invariant to translations and rotations, is also parity invariant, an obvious candidate for the set $\{\Omega ^\alpha \}$ is the parity operator $\pi $, since it commutes with the operators  $\mathbf J$ and $G$. It will be shown here that under the conditions of the theorem which are summarized at the end of this section, the matrix elements of the commutator $[G, H]$ are equal to zero on the subspace of the Hilbert space. This subspace is generated by the simultaneous eigenvectors of the set of commuting operators $\{H$, $\mathbf {J}^{2}$, $J_{z}$, $\{\Omega ^\alpha\}\}$ which belong to the Hilbert space. 

\subsection{Subspace $\mathcal {D}(H, \mathbf {J}^{2}, J_{z}, \{\Omega ^\alpha \})$}

Simultaneous eigenvectors of the set of commuting operators $\{H$, $\mathbf {J}^{2}$, $J_{z}$, $\{\Omega ^\alpha\}\}$ which belong to the Hilbert space are denoted
\begin{equation}
\label {eq:notationEsHJ2Jzpi} \vert n \rangle \equiv  \vert E_{n}, j _{n}, m _{n}, \{\Omega ^\alpha _n \} \rangle .
\end{equation}
In this simple notation $\vert n \rangle $ denotes the simultaneous eigenvector of operators $H$, $\mathbf {J}^{2}$, $J_{z}$ and $\{\Omega ^\alpha \}$, with the eigenvalues 
\begin{eqnarray}
\label{eq:env_n_} && H \vert n \rangle = E_{n}\vert n \rangle ,  \cr\nonumber\\
&& \mathbf {J}^{2}\vert n \rangle = j_{n}(j_{n}+1)\hbar ^{2}\vert n \rangle , \cr\nonumber\\
&& J_{z}\vert n \rangle = m_{n}\hbar \vert n \rangle , \cr\nonumber\\
&& \Omega ^\alpha  \vert n \rangle = \Omega ^\alpha _n\vert n \rangle , \qquad \qquad \forall \Omega ^\alpha \in \{\Omega ^\alpha \} .
\end{eqnarray}
If all members $\Omega ^\alpha$ of the set of operators $\{\Omega ^\alpha \}$ are known, the set of commuting operators $\{H$, $\mathbf {J}^{2}$, $J_{z}$, $\{\Omega ^\alpha\}\}$ is maximal. All members of the set of commuting operators $\{H$, $\mathbf {J}^{2}$, $J_{z}$, $\{\Omega ^\alpha\}\}$ are self-adjoint operators. Any simultaneous eigenvector of the set of commuting operators $\{H$, $\mathbf {J}^{2}$, $J_{z}$, $\{\Omega ^\alpha\}\}$, denoted by (\ref{eq:notationEsHJ2Jzpi}), is uniquely determined by the eigenvalues (\ref{eq:env_n_}). A basis in the subspace of the Hilbert space formed from the normalized simultaneous eigenvectors (\ref{eq:notationEsHJ2Jzpi}) is orthonormal
\begin{equation}
\langle n \vert m \rangle = \delta _{n, m} .
\end{equation}
From this point forward, this basis will be simply denoted by $\{\vert n \rangle \}$. Subspace of the Hilbert space generated by the basis $\{\vert n \rangle \}$ will be denoted $\mathcal {D}(H, \mathbf {J}^{2}, J_{z}, \{\Omega ^\alpha \})$.

The commutators of the operators $\mathbf {J}^{2}, \ J_{z}$ and $\{\Omega ^\alpha \}$ with the operator $G$, if taken between any $\vert n \rangle, \vert m \rangle \in \{\vert n \rangle \}$, are equal to:
\begin{eqnarray}
\label {eq:GcommJp}&& \langle n \vert [G, \mathbf {J}^{2}] \vert m \rangle =  \langle n \vert G \vert m \rangle [j_{m}(j_{m}+1) - j_{n}(j_{n}+1)]\hbar ^{2} = 0 , \cr\nonumber\\
&& \langle n \vert [G, J_{z}] \vert m \rangle =  \langle n \vert G \vert m \rangle (m_{m} - m _{n})\hbar = 0 , \cr\nonumber\\
&& \langle n \vert [G, \Omega ^\alpha ] \vert m \rangle =  \langle n \vert G \vert m \rangle (\Omega ^\alpha _m - \Omega ^\alpha _n) = 0, \qquad \qquad \forall \Omega ^\alpha \in \{\Omega ^\alpha \} . 
\end{eqnarray}
From the commutators (\ref {eq:GcommJpp}) and (\ref {eq:GcommOmega}), it follows that the matrix elements in (\ref {eq:GcommJp}) are equal to zero. Furthermore, from the matrix elements in (\ref {eq:GcommJp}), it follows that the matrix elements of the operator $G$ are equal to zero if taken between the eigenvectors with different $j$, between the eigenvectors with different $m$ and between the eigenvectors with different $\Omega ^\alpha $. This follows independently in all aforementioned cases, 
\begin{eqnarray}
\label {eq:expGcs} \langle n \vert G \vert m \rangle = \left \{ \begin{array}{l@{\,,\ }l } 
0 & \quad \forall \vert n \rangle , \vert m \rangle  \in \{\vert n \rangle \} : \ j_{n} \neq j_{m} \\
0 & \quad \forall \vert n \rangle , \vert m \rangle \in \{\vert n \rangle \} : \ m_{n} \neq m_{m} \\ 
0 & \quad \forall \vert n \rangle , \vert m \rangle \in \{\vert n \rangle \} : \ \Omega ^\alpha _n \neq \Omega ^\alpha _m \end{array} \right .. 
\end{eqnarray}
From the relation (\ref {eq:expGcs}), it follows that in aforementioned cases matrix elements of the commutator $[G, H]$ are equal to zero, i.e.
\begin{eqnarray}
\label {eq:expGcommH} \langle n \vert [G, H] \vert m \rangle = \langle n \vert G \vert m \rangle \left (E_m - E_n \right ) = \left \{ \begin{array}{l@{\,,\ }l} 
0 & \ \forall \vert n \rangle , \vert m \rangle \in \{\vert n \rangle \} : \ j_{n} \neq j_{m} \\
0 & \ \forall \vert n \rangle , \vert m \rangle \in \{\vert n \rangle \} : \ m_{n} \neq m_{m} \\ 
0 & \ \forall \vert n \rangle , \vert m \rangle \in \{\vert n \rangle \} : \ \Omega ^\alpha _n \neq \Omega ^\alpha _m \\
0 & \ \forall \vert n \rangle , \vert m \rangle \in \{\vert n \rangle \} : \ E _n = E _m \end{array} \right ..
\end{eqnarray}

It is clear, from relation (\ref {eq:expGcommH}), that to complete the proof of the relation $\langle n \vert [G, H] \vert m \rangle = 0$ for all eigenvectors $\vert n \rangle $ and $\vert m \rangle$, it still remains to show that the relation $ \langle n \vert G \vert m \rangle = 0$ is true for all cases (for all eigenvectors) not already contained in relation (\ref {eq:expGcs}) for which $E_{n} \neq E_{m}$. Therefore to complete the proof, it remains to show that the relation $ \langle n \vert G \vert m \rangle = 0$ is true for the eigenvectors  $\vert n \rangle $ and $\vert m \rangle$ for which $j_{n} = j_{m}, \ m_{n} =  m_{m}, \ \Omega ^\alpha_{n} = \Omega ^\alpha_{m}$ and $E_{n} \neq E_{m}$. Relation $\langle n \vert [G, H] \vert m \rangle = 0$ is established in this way by direct calculation, for all $\vert n \rangle, \vert m \rangle \in \{\vert n \rangle \}$, i.e. on the subspace $\mathcal {D}(H, \mathbf {J}^{2}, J_{z}, \{\Omega ^\alpha \})$ of the Hilbert space. 
\begin{pr}\label{proposition1}
Sufficient condition under which relations (\ref {eq:GcommJp}),  (\ref {eq:expGcs})  and (\ref {eq:expGcommH}) are defined without ambiguity, is that operator $G$ is defined on the subspace $\mathcal {D}(H, \mathbf {J}^{2}, J_{z}, \{\Omega ^\alpha \})$.\footnote{If $A$ and $B$ are both unbounded self-adjoint operators, the commutator $[A, B]$ a priori is only defined as a quadratic form on the intersection of their domains, i.e. on $D(A)\cap D(B)$ \cite{5}.} 
\end{pr}
Therefore, assumption that $\mathcal {D}(H, \mathbf {J}^{2}, J_{z}, \{\Omega ^\alpha \}) \subset D(G)$, where $D(G)$ represents the domain of operator $G$, should be tested initially.

\subsection{Analytic functions $f(\mathbf p)$}

The commutator of the operator $G$ with an analytic function $f(\mathbf {p})$ is equal to 
\begin{equation} 
\label {eq:Gcomm_an_fp} [G, f(\mathbf {p}) ] = i\hbar  \mathbf {p}\cdot {\partial f(\mathbf {p}) \over \partial \mathbf {p}} .
\end{equation} 
In relation (\ref{eq:Gcomm_an_fp}), the set of $N$ momentum operators $(\mathbf {p}_{1}, \dots, \mathbf {p}_{N})$ is denoted by $\mathbf p$ and $\mathbf {p}\cdot {\partial  \over \partial \mathbf {p}}$ is a notation for the directional derivative $\sum _{i=1}^{N} \mathbf {p}_{i} \cdot {\partial  \over \partial \mathbf {p}_{i}}$. For analytic functions $f(\mathbf {p})$ relation (\ref{eq:Gcomm_an_fp}) is derived in reference \cite{1} using the canonical commutation relations. For analytic functions\footnote{A function $f$ is called holomorphic (analytic) at a point $\alpha \in \mathbb{C}^n$ if it is holomorphic in some neighbourhood of this point. According to the Cauchy-Riemann criterion, a function of several variables which is holomorphic at a point $\alpha $ is holomorphic with respect to each variable (if the values of the other variables are fixed). The converse proposition is also true: if, in a neighbourhood of some point, a function $f$ is holomorphic with respect to each variable separately, then it is holomorphic at this point (Hartogs' fundamental theorem) \cite{8}.} $f(\mathbf P) = f\left (\sum_{i=1}^{N}\mathbf {p}_{i}\right ) $, the commutator with the operator $G$ is equal to
\begin{equation} 
\label {eq:Gcomm_an_ftP} [G, f(\mathbf {P}) ] = i\hbar  \mathbf {P}\cdot {\partial f(\mathbf {P}) \over \partial \mathbf {P}} .
\end{equation} 
Relation (\ref {eq:Gcomm_an_ftP}) is obtained from relation (\ref {eq:Gcomm_an_fp}) using the definition of the total momentum operator $\mathbf {P}=\sum_{i=1}^{N}\mathbf {p}_{i}$. 

It easy to show that the second commutator in (\ref {eq:GcommJpp}) can be obtained from the relation (\ref{eq:Gcomm_an_ftP}), if it is applied on the components $P_i$ of vector operator $\mathbf P$, where $i = 1,2,3$ ,  in place of the function $f(\mathbf {P})$. If the relation (\ref{eq:Gcomm_an_ftP}) is applied on the set of operators formed by the components $P_i$ of $r$ vector operators $\mathbf P$, 
\begin{equation}
\label {eq:PiPj...Pn} \underbrace{P_iP_j\cdots P_n}_{\mathrm{r \ operators}} , \ \qquad \underbrace{i,j,\dots, n}_{\mathrm{r \ indices}} = 1,2,3\ ,  
\end{equation}
it is straightforward to obtain
\begin{eqnarray}
\label {eq:G_comm_PiPj...Pn} [G, P_iP_j\cdots P_n] = i\hbar r P_iP_j\cdots P_n , \qquad i,j,\dots, n = 1,2,3\ .
\end{eqnarray}
Relation (\ref{eq:G_comm_PiPj...Pn}) applies for all $r$ where $r = 1,2,\dots$ . The operators (\ref {eq:PiPj...Pn}) can be considered components of the $3^r$-component tensor operator. The second commutator in (\ref {eq:GcommJpp}) represents a special case of (\ref{eq:G_comm_PiPj...Pn}) for $r=1$. 

The next step in the proof involves translational invariance of the Hamiltonian. The commutator (\ref {eq:HcommJp}) of the Hamiltonian $H$ with the generator of translations, operator of total momentum $\mathbf {P}$, is equal to zero. Using this commutation relation, one easily obtains that the commutators of $H$ with the operators (\ref {eq:PiPj...Pn}) are equal to zero,
\begin{eqnarray}
\label {eq:H_comm_PiPj...Pn} [H, P_iP_j\cdots P_n] = 0 , \ \qquad i,j,\dots, n = 1,2,3\ ,
\end{eqnarray}
for all $r  = 1,2,\dots \ $. For the matrix elements of the commutators (\ref {eq:H_comm_PiPj...Pn}) taken between any $\vert n \rangle, \vert m \rangle \in \{\vert n \rangle \}$, the following relation is obtained:
\begin{equation}
\label {eq:Hcommp_ME} \langle n  \vert [H, P_iP_j\cdots P_n] \vert m \rangle = \langle n \vert P_iP_j\cdots P_n \vert m \rangle (E_{n} - E_{m}) = 0 , 
\end{equation} 
for all $r = 1,2,\dots$ and $i,j,\dots, n = 1,2,3$. Relation (\ref{eq:Hcommp_ME}) gives
\begin{eqnarray} 
\label {eq:MEpnm} \langle n \vert P_iP_j\cdots P_n \vert m \rangle = 0 , & \qquad \forall \vert n \rangle , \vert m \rangle \in \{\vert n \rangle \} : \ E _n \neq E _m ,
\end{eqnarray}  
for all $r = 1,2,\dots$ and $i,j,\dots, n = 1,2,3$. 
\begin{pr}\label{proposition2}
Sufficient condition under which relations (\ref {eq:Hcommp_ME}) and (\ref {eq:MEpnm}) are defined without ambiguity, is that operators $P_iP_j\cdots P_n$ in (\ref {eq:PiPj...Pn}) are defined on the subspace $\mathcal {D}(H, \mathbf {J}^{2}, J_{z}, \{\Omega ^\alpha \})$, for all $r = 1,2,\dots$ and $i,j,\dots, n = 1,2,3$.
\end{pr}

\subsection{Projection operator into $\mathcal {D}(H, \mathbf {J}^{2}, J_{z}, \{\Omega ^\alpha \})$}

Any $\vert n \rangle \in \{\vert n \rangle \}$ is an eigenvector of the Hamiltonian $H$. Since the commutators (\ref {eq:H_comm_PiPj...Pn}) of the operators $P_iP_j\cdots P_n$ in (\ref {eq:PiPj...Pn}) with the Hamiltonian $H$ are equal to zero, it follows that $P_iP_j\cdots P_n \vert n \rangle $ is an eigenvector of $H$, or a zero vector. In either case
\begin{eqnarray}
\label{eq:ev_H_Pn} HP_iP_j\cdots P_n \vert n \rangle = E_n P_iP_j\cdots P_n \vert n \rangle . 
\end{eqnarray}
Relation (\ref{eq:ev_H_Pn}) follows in this way for all operators $P_iP_j\cdots P_n$ in (\ref {eq:PiPj...Pn}). The commutators of the operator $G$ with the operators $\mathbf J^2$, $J_{z}$ and $\{\Omega ^\alpha \}$ are equal to zero. It follows, for any $\vert n \rangle \in \{\vert n \rangle \}$, that $G \vert n \rangle $ is a simultaneous eigenvector of operators $\mathbf J^2$, $J_{z}$ and $\{\Omega ^\alpha \}$, or a zero vector. In either case
\begin{eqnarray}
\label{eq:ev_J2_Jz_Oa_Gn} && \mathbf J^2G\vert n \rangle = j_n(j_n +1)\hbar^2  G\vert n \rangle , \cr\nonumber\\
&& J_zG\vert n \rangle = m_n\hbar  G\vert n \rangle , \cr\nonumber\\
&& \Omega ^\alpha G\vert n \rangle = \Omega ^\alpha_n G\vert n \rangle , \qquad \qquad \forall \Omega ^\alpha \in \{\Omega ^\alpha \} .
\end{eqnarray}

Projection operator $\mathcal {P}_{\mathcal {D}(H)}$ maps every state vector onto its orthogonal projection in the subspace $\mathcal {D}(H)$, generated by the eigenvectors of the Hamiltonian $H$. Projection operator $\mathcal {P}_{\mathcal {D}(H)}$ is a bounded hermitian operator and it is idempotent, $\mathcal {P}_{\mathcal {D}(H)}^2 = \mathcal {P}_{\mathcal {D}(H)} $. This is true for projection operators into any subspace of Hilbert space \cite[348]{10}. Projection operator into the subspace $\mathcal {D}(\mathbf J^2, J_{z}, \{\Omega ^\alpha \})$, generated by the simultaneous eigenvectors of operators $\mathbf J^2$, $J_{z}$ and $\{\Omega ^\alpha \}$, is denoted by $\mathcal {P}_{\mathcal {D}(J^2, J_{z}, \{\Omega ^\alpha \})}$. Since the operators $H$, $\mathbf J^2$, $J_{z}$ and $\{\Omega ^\alpha \}$  form the set of commuting operators, it is straightforward to show that the projection operators $\mathcal {P}_{\mathcal {D}(J^2, J_{z}, \{\Omega ^\alpha \})}$ and $\mathcal {P}_{\mathcal {D}(H)}$ commute and
\begin{equation}
\label{eq:Proj_H_Proj_J2_Jz_Oa} \mathcal {P}_{\mathcal {D}(H)}\mathcal {P}_{\mathcal {D}(J^2, J_{z}, \{\Omega ^\alpha \})} = \mathcal {P}_{\mathcal {D}(J^2, J_{z}, \{\Omega ^\alpha \})}\mathcal {P}_{\mathcal {D}(H)} = \mathcal {P}_{\mathcal {D}(H, J^2, J_{z}, \{\Omega ^\alpha \})} .
\end{equation}

Introduced here in relation (\ref{eq:Proj_H_Proj_J2_Jz_Oa}), $\mathcal {P}_{\mathcal {D}(H, J^2, J_{z}, \{\Omega ^\alpha \})}$ is a projection operator into the subspace $\mathcal {D}(H, \mathbf {J}^{2}, J_{z}, \{\Omega ^\alpha \})$ generated by the basis $\{\vert n \rangle \}$ of simultaneous eigenvectors of the set of commuting operators $\{H$, $\mathbf {J}^{2}$, $J_{z}$, $\{\Omega ^\alpha\}\}$. Relation (\ref{eq:Proj_H_Proj_J2_Jz_Oa}) is a special case of the property of projection operators associated to self-adjoint commuting operators. It follows from the definition of commutativity for self-adjoint operators and the properties of their associated projection-valued measures, \cite[Vol. 1]{4}. It is easily seen from relation (\ref{eq:Proj_H_Proj_J2_Jz_Oa}), that $\mathcal {P}_{\mathcal {D}(H, J^2, J_{z}, \{\Omega ^\alpha \})}$ is hermitian and idempotent. With the help of relations (\ref{eq:ev_H_Pn}), (\ref{eq:ev_J2_Jz_Oa_Gn}) and  (\ref{eq:Proj_H_Proj_J2_Jz_Oa}), the following relations are obtained,
\begin{equation}
\label{eq:ME_n_m_G_Proj_H_j2_Jz_Oa_P} \langle n \vert G P_iP_j\cdots P_n   \vert m \rangle = \langle n \vert G\mathcal {P}_{\mathcal {D}(H, J^2, J_{z}, \{\Omega ^\alpha \})} P_iP_j\cdots P_n \vert m \rangle ,
\end{equation}
and
\begin{equation}
\label{eq:ME_n_m_P_Proj_H_j2_Jz_Oa_G} \langle n \vert P_iP_j\cdots P_n G \vert m \rangle = \langle n \vert P_iP_j\cdots P_n \mathcal {P}_{\mathcal {D}(H, J^2, J_{z}, \{\Omega ^\alpha \})} G \vert m \rangle ,
\end{equation}
for all operators $P_iP_j\cdots P_n$ given by (\ref {eq:PiPj...Pn}) and for all $\vert n \rangle, \vert m \rangle \in \{\vert n \rangle \}$.

From relation (\ref {eq:G_comm_PiPj...Pn}) it follows that on the subspace $\mathcal {D}(H, \mathbf {J}^{2}, J_{z}, \{\Omega ^\alpha \})$ the expectation value of the operator $P_iP_j\cdots P_n$ is equal (up to a constant factor) to the expectation value of the commutator $[G,  P_iP_j\cdots P_n]$. This inference is correct under conditions introduced in the analysis in previous subsections. It is useful when calculating matrix elements of the operators $P_iP_j\cdots P_n$ in (\ref {eq:PiPj...Pn}). For the expectation value taken with respect to any $\vert n \rangle \in \{\vert n \rangle \}$, relation (\ref {eq:G_comm_PiPj...Pn}) gives 
\begin{equation} 
\label {eq:ME_GcommP} \langle n \vert [G, P_iP_j\cdots P_n] \vert n \rangle = i\hbar r \langle n \vert P_iP_j\cdots P_n \vert n \rangle .
\end{equation} 
With the help of relations (\ref{eq:ME_n_m_G_Proj_H_j2_Jz_Oa_P}) and (\ref{eq:ME_n_m_P_Proj_H_j2_Jz_Oa_G}), it is easy to show that the left side of relation (\ref {eq:ME_GcommP}) is equal to 
\begin{equation} 
\label {eq:ME_GcommP_lhs} \langle n \vert G\mathcal{P}_{\mathcal {D}(H, J^{2}, J_{z}, \{\Omega ^\alpha \})}P_iP_j\cdots P_n \vert n \rangle - \langle n \vert P_iP_j\cdots P_n \mathcal{P}_{\mathcal {D}(H, J^{2}, J_{z}, \{\Omega ^\alpha \} )} G \vert n \rangle . 
\end{equation}
Then, using relations (\ref{eq:expGcs}), (\ref {eq:MEpnm}) and (\ref {eq:ME_GcommP_lhs}) one obtains 
\begin{equation}
\label{eq:Gpdiff} \langle n \vert G \vert n \rangle \left \{\langle n \vert P_iP_j\cdots P_n \vert n \rangle - \langle n \vert P_iP_j\cdots P_n \vert n \rangle \right \} = i\hbar r \langle n \vert P_iP_j\cdots P_n \vert n \rangle .
\end{equation}
The factor in the brackets on the left side of relation (\ref{eq:Gpdiff}) is equal to zero. It follows from relation (\ref{eq:Gpdiff}) that the expectation values of operators $P_iP_j\cdots P_n$ in (\ref {eq:PiPj...Pn}), taken with respect to all $\vert n \rangle \in \{\vert n \rangle \}$ are equal to zero
\begin{eqnarray}
\label {eq:n_PiPj...Pn_n_0}  \langle n \vert P_iP_j\cdots P_n \vert n \rangle = 0 , & \qquad \forall \vert n \rangle \in \{\vert n\rangle \} \subset \mathcal {D}(H, \mathbf {J}^{2}, J_{z}, \{\Omega ^\alpha \}) .
\end{eqnarray}
Relation (\ref{eq:n_PiPj...Pn_n_0}) is true for all operators $P_iP_j\cdots P_n$ in (\ref {eq:PiPj...Pn}). By applying similar procedures when calculating the matrix elements of commutators (\ref {eq:G_comm_PiPj...Pn}) between  all $\vert n \rangle, \vert m \rangle \in \{\vert n \rangle \}$, and using relations (\ref{eq:expGcs}), (\ref{eq:MEpnm}) and (\ref{eq:n_PiPj...Pn_n_0}) one obtains
\begin{eqnarray}
\label {eq:n_PiPj...Pn_m_0}  \langle n \vert P_iP_j\cdots P_n \vert m \rangle = 0 , & \qquad \forall \vert n \rangle , \vert m \rangle \in \{\vert n\rangle \} \subset \mathcal {D}(H, \mathbf {J}^{2}, J_{z}, \{\Omega ^\alpha \}) .
\end{eqnarray}
Under sufficient conditions, required by Definition \ref{definition1} and Propositions \ref{proposition1} and \ref{proposition2}, relation (\ref{eq:n_PiPj...Pn_m_0}) is true for all operators $P_iP_j\cdots P_n$ in (\ref {eq:PiPj...Pn}).

\subsection{Multiple power series} 

Analytic function $f(\mathbf {P})$ is represented by a multiple power series. In abbreviated notation it is written as
\begin{equation}
\label {eq:p_s_e_f_P} f(\mathbf {P}) = \sum_{k=0}^{\infty }\frac{1}{k!}\frac{\partial ^{k}f}{\partial \mathbf {P}^{k}}(0) \mathbf {P}^{k} ,
\end{equation} 
with the terms in the series (\ref{eq:p_s_e_f_P}) given by
\begin{equation}
\label{eq:tmpsfP} \frac{1}{k!}\frac{\partial ^{k}f}{\partial \mathbf {P}^{k}}(0) \mathbf {P}^{k} \equiv  \sum _{k_1, k_2, k_3 \geq 0}^{k_1 + k_2 + k_3 = k} \frac{1}{k_1!k_2! k_3!}\frac{\partial ^{k_1 + k_2 + k_3 }f}{\partial P_{1}^{k_1} \partial P_{2}^{k_2}\partial P_{3}^{k_3}}(0)P_{1}^{k_1} P_{2}^{k_2}P_{3}^{k_3} .
\end{equation}
It is clear from (\ref{eq:tmpsfP}) that all terms in the multiple power series (\ref {eq:p_s_e_f_P}) are equal (up to constant factors) to operators $P_{1}^{k_1} P_{2}^{k_2}P_{3}^{k_3}$, which are of the form (\ref {eq:PiPj...Pn}):
\begin{equation}
\label {eq:Pk_PiPj...Pn} P_{1}^{k_1} P_{2}^{k_2}P_{3}^{k_3} = \underbrace{P_1\cdots P_1}_{\mathrm{k_1}} \underbrace{P_2\cdots P_2}_{\mathrm{k_2}} \underbrace{P_3\cdots P_3}_{\mathrm{k_3}} .
\end{equation}
Relation (\ref{eq:n_PiPj...Pn_m_0}) is useful for evaluating matrix elements of the function (\ref {eq:p_s_e_f_P}) between any $\vert n \rangle, \vert m \rangle \in \{\vert n \rangle \}$.  As it clearly follows from relations (\ref {eq:n_PiPj...Pn_m_0}), (\ref{eq:tmpsfP}) and (\ref {eq:Pk_PiPj...Pn}), the matrix elements of the function $f(\mathbf {P})$ represented by (\ref {eq:p_s_e_f_P}) between all $\vert n \rangle, \vert m \rangle \in \{\vert n \rangle \}$ are equal to
\begin{equation} 
\label {eq:f_P_p_s_e_En_Em} \langle n \vert f(\mathbf {P}) \vert m \rangle = f(0) \langle n \vert  m \rangle = f(0) \delta _{n, m} .
\end{equation}
If the coefficient $f(0)$ in the multiple power series (\ref {eq:p_s_e_f_P}) is equal to zero, it follows that all matrix elements in (\ref{eq:f_P_p_s_e_En_Em}) are equal to zero. Under conditions in this analysis, relation (\ref{eq:f_P_p_s_e_En_Em}) is true for all functions $f(\mathbf {P})$ represented by multiple power series (\ref {eq:p_s_e_f_P}).

\subsection{Translational invariance in $\mathcal {D}(H, \mathbf {J}^{2}, J_{z}, \{\Omega ^\alpha \})$}

{\sloppy The translation operator for a spatial displacement of the system by a displacement vector $\mathbf d$ is given by 
\begin{equation}
\label {eq:TO_d} U( \mathbf d ) = \exp \left ( \frac{-i\mathbf d \cdot \mathbf P}{\hbar }\right ) ,
\end{equation}
Total momentum operator $\mathbf P = \sum_{i=1}^{N} \mathbf p$ divided by $\hbar$ is the generator of translations for the system. Translation operator $U( \mathbf d )$ given in (\ref {eq:TO_d}) is represented by a multiple power series of the form (\ref {eq:p_s_e_f_P}). From relation (\ref{eq:f_P_p_s_e_En_Em}) applied on the translation operator $U( \mathbf d )$, one obtains
\begin{equation} 
\label {eq:ME_n_m_TO_d} \langle n \vert U( \mathbf d ) \vert m \rangle = \langle n \vert  m \rangle = \delta _{n, m},
\end{equation}
for all $\vert n \rangle, \vert m \rangle \in \{\vert n \rangle \}$. Translation operator $U( \mathbf d )$ is a unitary operator, $U( \mathbf d )^\dagger U( \mathbf d ) =  U( \mathbf d )U( \mathbf d )^\dagger  = I$. From the unitarity of $U( \mathbf d )$ it follows that 
\begin{equation} 
\label{eq:n_n_TO_d_Un} \langle n \vert U( \mathbf d )^\dagger U( \mathbf d ) \vert n \rangle = \langle n \vert U( \mathbf d )U( \mathbf d )^\dagger  \vert n \rangle = \langle n \vert  n \rangle = 1 .
\end{equation}
Relations (\ref{eq:ME_n_m_TO_d}) and (\ref{eq:n_n_TO_d_Un}) are equivalent to invariance of all $\vert n \rangle \in \{\vert n \rangle \}$ to translations,
\begin{equation} 
\label {eq:TO_d_n} U( \mathbf d ) \vert n \rangle = \vert  n \rangle .
\end{equation}

}

Eigenfunctions $\psi _n$ correspond to simultaneous eigenvectors $\vert n \rangle $ of the set of commuting operators $\{H$, $\mathbf {J}^{2}$, $J_{z}$, $\{\Omega ^\alpha\}\}$. The basis of simultaneous eigenvectors $\vert \mathbf r_1, \dots, \mathbf r_N \rangle$ of the set of $N$ position operators $(\mathbf r_1, \dots , \mathbf r_N)$ is introduced in the Appendix \ref{app:1}. In the usual notation of wave mechanics eigenfunctions $\psi _n$ are recognized as
\begin{equation}
\label {eq:ev_n_p_b_v} \langle \mathbf r_1, \dots, \mathbf r_N \vert n \rangle = \psi _n (\mathbf r_1 , \dots, \mathbf r_N) .
\end{equation}
Equivalently, relation (\ref {eq:TO_d_n}) implies that the eigenfunctions (\ref {eq:ev_n_p_b_v}) are invariant to space displacements and depend on the relative position vectors of $N$ particles, 
\begin{equation}
\label {eq:TO_d_psi_n} \psi _n(\mathbf r_1, \dots, \mathbf r_N) = \psi _n(\{\mathbf r_i - \mathbf r_j\}) .
\end{equation}
In Eq. (\ref {eq:TO_d_psi_n}), $\{\mathbf r_i - \mathbf r_j\}$ is the set of relative position vectors $\mathbf r_i - \mathbf r_j$, where $i,j = 1, \dots , N$ and $i\neq j$. Furthermore, Eqs. (\ref{eq:ev_n_p_b_v}) and (\ref {eq:TO_d_psi_n}) imply that for all operators $P_iP_j\cdots P_n$ in (\ref {eq:PiPj...Pn}), and for all $\vert n \rangle \in \{\vert n \rangle \}$, 
\begin{eqnarray}
\label {eq:PiPj_Pn_n_eq_0} P_iP_j\cdots P_n \vert n \rangle = 0 , & \qquad \forall \vert n \rangle \in \{\vert n\rangle \} \subset \mathcal {D}(H, \mathbf {J}^{2}, J_{z}, \{\Omega ^\alpha \}) .
\end{eqnarray}
Relation (\ref {eq:PiPj_Pn_n_eq_0}) then implies
\begin{eqnarray} 
f(\mathbf {P}) \vert n \rangle = f(0) \vert n \rangle , & \qquad \forall \vert n \rangle \in \{\vert n\rangle \} \subset \mathcal {D}(H, \mathbf {J}^{2}, J_{z}, \{\Omega ^\alpha \}) ,\label {eq:f_P_n_eq_0}
\end{eqnarray}
for all functions $f(\mathbf {P})$ represented by multiple power series (\ref {eq:p_s_e_f_P}).

\subsection{Gaussian function $\phi _a(\mathbf P, \mathbf K)$}

Important function represented by a multiple power series of the form (\ref {eq:p_s_e_f_P}) is the Gaussian function
\begin{equation}
\label{eq:Gauss_P_3D} \phi _a(\mathbf P, \mathbf K) = \frac{1}{\left (2a\sqrt{\pi} \right )^3}\exp\left [-\frac{(\mathbf P -\mathbf K )^2}{4a^2}\right ] ,
\end{equation}
where $\mathbf P$ is the total momentum operator and $\mathbf K$ is the vector in Euclidean space $\mathbb{R}^3$. It is known that the Gaussian function $\phi _a(\mathbf x )$ for $\mathbf x \in \mathbb{R}^3$ corresponds to the delta distribution $\delta ^3(\mathbf x)$ in the sense explained for example in reference \cite{11}. Appropriate correspondence is investigated for the function $\phi _a (\mathbf P, \mathbf K)$ of the total momentum operator $\mathbf P$ in the next subsection and in the Appendix \ref{app:1}. From the relation (\ref {eq:f_P_p_s_e_En_Em}) for the functions represented by a multiple power series (\ref {eq:p_s_e_f_P}), it follows that the matrix elements of the Gaussian function $\phi _a (\mathbf P, \mathbf K)$, taken between all $\vert n \rangle, \vert m \rangle \in \{\vert n \rangle \}$ are equal to 
\begin{equation}
\label{eq:E_Gauss_n} \langle n \vert \phi _a (\mathbf P, \mathbf K ) \vert m \rangle  = \phi _a (0, \mathbf K) \langle n \vert m \rangle = \left (2a\sqrt{\pi} \right )^{-3}e^{-\frac{\mathbf K ^2}{4a^2}} \delta _{n, m} .
\end{equation}

\subsection{Operator $\phi _a(\mathbf P, \mathbf K)$ and $a \rightarrow 0$ limit}

{\sloppy Provided that the limit $a \rightarrow 0$ of the matrix element of the function $\phi _a(\mathbf P, \mathbf K)$ taken between the vectors $\vert \Psi \rangle $ and $\vert \Phi \rangle $ in the Hilbert space is finite, this limit is denoted as
\begin{equation}
\label{eq:lim_Gauss_Sv_Sv} \lim_{a \rightarrow 0} \langle \Phi  \vert \phi _a (\mathbf P , \mathbf K) \vert \Psi \rangle  = \langle \Phi \vert \delta ^3 (\mathbf P, \mathbf K) \vert \Psi  \rangle .
\end{equation}
The limit (\ref{eq:lim_Gauss_Sv_Sv}) is evaluated in relations (\ref{eq:ps_psp_Gauss_lim}), (\ref{eq:ps_psp_Gauss_delta_lim}) and (\ref{eq:ps_psp_delta_P}) in the Appendix \ref{app:1}. It is shown in the Appendix \ref{app:1}, that the matrix representation $\langle \mathbf p \vert \delta ^3(\mathbf P, \mathbf K) \vert \mathbf p^\prime \rangle $ in the basis $\vert \mathbf p_1, \dots, \mathbf p_N \rangle$ in relation (\ref{eq:ME_p_pp_delta_P}), is related to the matrix representation in the basis $\vert \mathbf r_1, \dots, \mathbf r_N \rangle$ by a unitary transformation. In this way the matrix $\langle \mathbf r \vert \delta ^3(\mathbf P, \mathbf K) \vert \mathbf r^\prime \rangle $ in relation (\ref{eq:ME_r_rp_delta_P}) is obtained. With the inverse transformation the matrix $\langle \mathbf p \vert \delta ^3(\mathbf P, \mathbf K) \vert \mathbf p^\prime \rangle $ is again obtained. This is important as the matrix representations (\ref{eq:ME_p_pp_delta_P}) and (\ref{eq:ME_r_rp_delta_P}) are useful in evaluating the limit (\ref{eq:lim_Gauss_Sv_Sv}). However, it should be stressed for clarity that neither vectors $\vert \mathbf p_1, \dots, \mathbf p_N \rangle $ nor vectors $\vert \mathbf r_1, \dots, \mathbf r_N \rangle $ introduced in the Appendix \ref{app:1}, are normalizable and do not themselves actually belong to Hilbert space. Therefore, in the technical sense, they do not form a basis in the Hilbert space.

}

For the vectors $\vert \Psi \rangle $ and $\vert \Phi  \rangle $ in the Hilbert space, the limit (\ref{eq:lim_Gauss_Sv_Sv}) is evaluated here using the relation (\ref{eq:ME_r_rp_delta_P}). In the limit (\ref{eq:lim_Gauss_Sv_Sv}) one obtains
\begin{equation}
\label{eq:lim_int_Sv_Gauss_Sv} \langle \Phi \vert \delta ^3 (\mathbf P, \mathbf K) \vert \Psi \rangle = \left (2\pi \hbar\right)^{-3} \mathcal{I}_k(\mathbf K) , 
\end{equation}
where $\mathcal{I}_k(\mathbf K)$ is the integral
\begin{equation}
\label{eq:int_I_K} \mathcal{I}_k(\mathbf K) = \int  e^{\frac{i}{\hbar }\mathbf K \cdot \left (\mathbf r_k - \mathbf r_k^\prime \right ) }\Phi ^\ast (\mathbf r_1, \dots, \mathbf r_N) \Psi (\mathbf r_1 + \mathbf {r}_k^\prime  - \mathbf r_k, \dots ,\mathbf r_N + \mathbf {r}_k ^\prime - \mathbf r_k)d^3r^\prime _k\prod_{i=1}^Nd^3r_i .
\end{equation} 
Functions $\Psi (\mathbf r_1, \dots, \mathbf r_N)$ and $\Phi (\mathbf r_1, \dots, \mathbf r_N)$ correspond to vectors  $\vert \Psi \rangle $ and $\vert \Phi \rangle $ in the Hilbert space. The vector $\mathbf {r}^\prime_k$ in relation (\ref{eq:int_I_K}) is any position vector from the set ($\mathbf {r}^\prime_1, \dots , \mathbf {r}^\prime_N $). If the functions $\Psi (\mathbf r_1, \dots, \mathbf r_N)$ and $\Phi (\mathbf r_1, \dots, \mathbf r_N)$ belong to the vector space $D(\mathbb{R}^{3N})$ of infinitely differentiable functions on $\mathbb{R}^{3N}$ with compact support, the integral (\ref{eq:int_I_K}) is convergent and the limit (\ref{eq:lim_Gauss_Sv_Sv}) is finite. All functions in $D(\mathbb{R}^{3N})$ belong to the Hilbert space  $L^2(\mathbb{R}^{3N})$ of square integrable complex-valued functions and $D(\mathbb{R}^{3N})$ is a dense subspace of $L^2(\mathbb{R}^{3N})$.

The limit (\ref{eq:lim_Gauss_Sv_Sv}) is defined for all $\vert n \rangle, \vert m \rangle \in \{\vert n \rangle \}$. The eigenfunctions $\psi _n$ and $\psi _m$ correspond to eigenvectors $\vert n \rangle $ and $\vert m \rangle $, as given by the relation (\ref {eq:ev_n_p_b_v}). Using the property of translational invariance (\ref {eq:TO_d_psi_n}) of eigenfunctions $\psi _n $, one obtains the following equality from relations (\ref{eq:lim_Gauss_Sv_Sv}), (\ref{eq:lim_int_Sv_Gauss_Sv}) and (\ref{eq:int_I_K}), 
\begin{equation}
\label{eq:lim_int_n_Gauss_m} \langle n \vert \delta ^3 (\mathbf P, \mathbf K  ) \vert m \rangle = \delta ^3(\mathbf K ) \int e^{\frac{i}{\hbar }\mathbf K \cdot \mathbf r_k} \psi _n ^\ast (\{\mathbf r_j - \mathbf r_l\}) \psi _m(\{\mathbf r_j - \mathbf r_l\})\prod_{i=1}^Nd^3r_i .
\end{equation}
Relation (\ref{eq:lim_int_n_Gauss_m}) can be written in the form
\begin{equation}
\label{eq:ME_dd_P_n_m} \langle n \vert \delta ^3 (\mathbf P, \mathbf K) \vert m \rangle = \delta ^3(\mathbf K ) \mathcal {F}_{(n,m) , k}(\mathbf K) .  
\end{equation}
The function $\mathcal {F}_{(n,m), k}(\mathbf K)$ in relation (\ref{eq:ME_dd_P_n_m}) is equal to the Fourier transform
\begin{equation}
\label{eq:F_T_fun_F} \mathcal {F}_{(n,m), k}(\mathbf K) = \int e^{\frac{i}{\hbar }\mathbf K \cdot \mathbf r_k} \psi _n ^\ast (\{\mathbf r_j - \mathbf r_l\}) \psi _m(\{\mathbf r_j - \mathbf r_l\})\prod_{i=1}^Nd^3r_i .
\end{equation}
The Fourier transform (\ref{eq:F_T_fun_F}) is defined since all eigenfunctions $\psi _n(\{\mathbf r_j - \mathbf r_l\})$ belong to the Hilbert space  $L^2(\mathbb{R}^{3N})$ of square integrable complex-valued functions. Therefore, functions\linebreak $e^{\frac{i}{\hbar }\mathbf K \cdot \mathbf r_k}\psi _n(\{\mathbf r_j - \mathbf r_l\})$ also belong to the Hilbert space $L^2(\mathbb{R}^{3N})$ and the integral (\ref{eq:F_T_fun_F}) is convergent. The following properties of $\mathcal {F}_{(n,m), k}(\mathbf K)$ are observed, 
\begin{equation}
\label{eq:F_n_m_k_Keq0} \mathcal {F}_{(n,m), k}(0) = \langle n \vert m \rangle = \delta _{n, m} ,  
\end{equation} 
and 
\begin{equation}
\label{eq:F_n_m_k_eq_F_n_m_l} \mathcal {F}_{(n,m), k}(0) = \mathcal {F}_{(n,m), l}(0) ,  
\end{equation} 
for all $k,l = 1, \dots, N$. The indices $(n,m)$ correspond to all $\vert n \rangle, \vert m \rangle \in \{\vert n \rangle \}$.

Relations (\ref{eq:ME_dd_P_n_m}), (\ref{eq:F_T_fun_F}), (\ref{eq:F_n_m_k_Keq0}) and (\ref{eq:F_n_m_k_eq_F_n_m_l}) are consistent with the limit $a \rightarrow 0$ of relation (\ref{eq:E_Gauss_n}). From relations (\ref{eq:E_Gauss_n}) and (\ref{eq:lim_Gauss_Sv_Sv}) it is clear that the matrix elements $\langle n \vert \delta ^3 (\mathbf P, \mathbf K = 0) \vert m \rangle$ between the eigenvectors $\{\vert n \rangle \}$ are unbounded. Relations (\ref{eq:lim_int_Sv_Gauss_Sv}), (\ref{eq:int_I_K}) and (\ref{eq:lim_int_n_Gauss_m}) indicate that operator $\delta ^3 (\mathbf P, \mathbf K=0)$ is not defined on the subspace $\mathcal {D}(H, \mathbf {J}^{2}, J_{z}, \{\Omega ^\alpha \})$ of the Hilbert space. This is verified with the help of  relations (\ref {eq:ev_n_p_b_v}), (\ref {eq:TO_d_psi_n}) and (\ref{eq:ME_r_rp_delta_P}),
\begin{equation}
\label{eq:norm_delta_n}  \| \delta ^3 (\mathbf P, \mathbf K  ) \vert n \rangle \|^{2} = \delta ^3(\mathbf K )\delta ^3(\mathbf K ) ,
\end{equation}
for all $\vert n \rangle \in \{\vert n \rangle \}$. The norm of a vector $\vert \Psi \rangle $ in the Hilbert space, denoted by $\| \vert \Psi \rangle \| $, is defined to be the nonnegative real number $\| \vert \Psi \rangle \|  = \sqrt{\langle \Psi \vert \Psi \rangle } \geq 0 $. In the representation (\ref{eq:ME_r_rp_delta_P}) it is straightforward to show that the vector subspace $D_{\partial_x} (\mathbb{R}^{3N})$ of the Hilbert space $L^2(\mathbb{R}^{3N})$,
\begin{equation}
D_{\partial_x}(\mathbb{R}^{3N}) = \left \{ \varphi  (\mathbf r_1, \dots, \mathbf r_N) \left | \ \varphi  = \sum_{i=1}^{N}\frac{\partial \chi }{\partial x_i} , \ \ \chi  (\mathbf r_1, \dots, \mathbf r_N) \in D(\mathbb{R}^{3N}) \right. \right \} ,
\end{equation}
is a vector subspace of the null space of operator $\delta ^3 (\mathbf P, \mathbf K=0)$  and also that $D_{\partial_x} (\mathbb{R}^{3N}) \subset D(\mathbb{R}^{3N})$. The same is true for $D_{\partial_y} (\mathbb{R}^{3N})$ and $D_{\partial_z} (\mathbb{R}^{3N})$.

\subsection{Properties of operator $G$ on $\mathcal {D}(H, \mathbf {J}^{2}, J_{z}, \{\Omega ^\alpha \})$}

The commutator of the operator $G$ with the function $\phi _a (\mathbf P, \mathbf K )$ follows from relation (\ref {eq:Gcomm_an_ftP}):
\begin{equation}
\label {eq:G_comm_Gauss} [G, \phi _a (\mathbf P, \mathbf K)] = i\hbar \mathbf P \cdot \frac{\partial \phi _a(\mathbf P,\mathbf K)}{\partial \mathbf P} . 
\end{equation}
With the help of relations (\ref {eq:f_P_p_s_e_En_Em}) and (\ref{eq:Gauss_P_3D}), one obtains that the matrix elements of the commutator (\ref {eq:G_comm_Gauss}) taken between all $\vert n \rangle, \vert m \rangle \in \{\vert n \rangle \}$ are equal to zero,
\begin{equation}
\label {eq:ME_n_m_G_comm_Gauss} \langle n \vert [G, \phi _a (\mathbf P, \mathbf K )] \vert m \rangle = 0 . 
\end{equation}
Function $\phi _a (\mathbf P, \mathbf K)$ is represented by a multiple power series of the form (\ref {eq:p_s_e_f_P}), and for any such function relation (\ref{eq:f_P_n_eq_0}) holds. Therefore, for any $\vert n \rangle \in \{\vert n \rangle \}$, it follows that $\phi _a (\mathbf P, \mathbf K ) \vert n \rangle $ is an eigenvector of $H$,
\begin{equation}
\label {eq:ev_H_Gauss_P_n} H\phi _a (\mathbf P, \mathbf K ) \vert n \rangle = E_n \phi _a (\mathbf P, \mathbf K) \vert n \rangle .
\end{equation}
With the help of relations (\ref{eq:ev_J2_Jz_Oa_Gn}), (\ref{eq:Proj_H_Proj_J2_Jz_Oa}) and (\ref{eq:ev_H_Gauss_P_n}), one obtains the following relations
\begin{equation}
\label {eq:ME_n_m_G_Proj_H_j2_Jz_Oa_Gauss_P} \langle n \vert G \phi _a (\mathbf P, \mathbf K)  \vert m \rangle = \langle n \vert G\mathcal {P}_{\mathcal {D}(H, J^2, J_{z}, \{\Omega ^\alpha \})} \phi _a (\mathbf P, \mathbf K) \vert m \rangle ,
\end{equation}
and
\begin{equation}
\label {eq:ME_n_m_Gauss_P_Proj_H_j2_Jz_Oa_G} \langle n \vert \phi _a (\mathbf P, \mathbf K) G \vert m \rangle = \langle n \vert \phi _a (\mathbf P, \mathbf K) \mathcal {P}_{\mathcal {D}(H, J^2, J_{z}, \{\Omega ^\alpha \})} G \vert m \rangle ,
\end{equation}
for all $\vert n \rangle, \vert m \rangle \in \{\vert n \rangle \}$. With the help of (\ref{eq:ME_n_m_G_Proj_H_j2_Jz_Oa_Gauss_P}) and (\ref{eq:ME_n_m_Gauss_P_Proj_H_j2_Jz_Oa_G}), relation (\ref {eq:ME_n_m_G_comm_Gauss}) becomes 
\begin{equation} 
\label {eq:ME_G_comm_Gauss_P} \langle n \vert G\mathcal{P}_{\mathcal {D}(H, J^{2}, J_{z}, \{\Omega ^\alpha \})}\phi _a (\mathbf P, \mathbf K)  \vert m \rangle - \langle n \vert \phi _a (\mathbf P, \mathbf K) \mathcal{P}_{\mathcal {D}(H, J^{2}, J_{z}, \{\Omega ^\alpha \} )} G \vert m \rangle = 0 . 
\end{equation}
Then, from relation (\ref {eq:ME_G_comm_Gauss_P}), with the help of relation (\ref{eq:E_Gauss_n}) one obtains 
\begin{equation}
\label {eq:ME_n_m_G_comm_Gauss_eq_0} \langle n \vert G \vert m \rangle \left (2a\sqrt{\pi} \right )^{-3}e^{-\frac{\mathbf K ^2}{4a^2}} - \left (2a\sqrt{\pi} \right )^{-3}e^{-\frac{\mathbf K ^2}{4a^2}} \langle n \vert G \vert m \rangle  = 0 .
\end{equation}

The notation introduced in the Appendix \ref{app:1}, in particular in relations (\ref{eq:ps_psp_Gauss_lim}), (\ref{eq:ps_psp_Gauss_delta_lim}) and (\ref{eq:ps_psp_delta_P}), suggests that it is possible to write
\begin{equation}
\label{eq:ps_psp_PdGauss_PdD_lim} \lim _{a\rightarrow 0} \langle \Phi \vert \mathbf P \cdot \frac{\partial \phi _a(\mathbf P, \mathbf K )}{\partial \mathbf P}  \vert \Psi \rangle = \langle \Phi  \vert \mathbf P \cdot \frac{\partial \delta ^3(\mathbf P, \mathbf K)}{\partial \mathbf P}  \vert \Psi \rangle , 
\end{equation}
where $\vert \Psi \rangle $ and $\vert \Phi  \rangle $ are vectors in the Hilbert space. In this way the following relation is obtained, 
\begin{equation}
\label{eq:ps_psp_PdP} \langle \Phi \vert \mathbf P \cdot \frac{\partial \delta ^3(\mathbf P, \mathbf K)}{\partial \mathbf P}  \vert \Psi \rangle = \int \langle \Phi \vert \mathbf p  \rangle \mathbf P \cdot \frac{\partial \delta ^3(\mathbf P - \mathbf K )}{\partial \mathbf P} \langle \mathbf p  \vert \Psi \rangle \prod _{i=1}^N d^3p_i .
\end{equation}
In relation (\ref{eq:ps_psp_PdP}) $\vert \mathbf p \rangle $ is a notation for $\vert \mathbf p_1, \dots, \mathbf p_N \rangle$. The functions next to the derivative of the delta distribution $\delta ^3(\mathbf P - \mathbf K)$ in the integral (\ref{eq:ps_psp_PdP}) may have some properties of test functions. If the functions $\langle \mathbf p  \vert \Psi \rangle$ and $\langle \mathbf p  \vert \Phi  \rangle $ do not belong to the space $D(\mathbb{R}^{3N})$ of infinitely differentiable functions on $\mathbb{R}^{3N}$ with compact support (the space of test functions), they still belong to the Hilbert space of square integrable functions $L^2(\mathbb{R}^{3N})$. Depending on differentiability of the functions $\langle \mathbf p  \vert \Psi \rangle$ and $\langle \mathbf p  \vert \Phi \rangle $, the rules of $\delta $-calculus for derivatives of delta distribution and partial integration may still be applicable.

It is straightforward to prove that for all $\vert n \rangle, \vert m \rangle \in \{\vert n \rangle \}$ the corresponding limit (\ref{eq:ps_psp_PdGauss_PdD_lim}) is finite. In a direct way, using the relation (\ref {eq:f_P_p_s_e_En_Em}) for the matrix elements of functions (\ref{eq:p_s_e_f_P}) between $\vert n \rangle, \vert m \rangle \in \{\vert n \rangle \}$, and the relation (\ref{eq:Gauss_P_3D}), one obtains 
\begin{equation}
\label{eq:n_m_PdP_Gauss_lim} \lim_{a \rightarrow 0} \langle n \vert \mathbf P \cdot \frac{\partial \phi _a(\mathbf P, \mathbf K )}{\partial \mathbf P}  \vert m \rangle  = \lim_{a \rightarrow 0} \left ( 0 \right ) = 0 .
\end{equation}
Therefore, for all $\vert n \rangle, \vert m \rangle \in \{\vert n \rangle \}$, the limit (\ref{eq:ps_psp_PdGauss_PdD_lim}) is finite and is equal to zero. The limiting procedure is formally equivalent to calculating the matrix elements of the commutator  
\begin{equation}
\label {eq:comm_G_delta3_P_K} [G, \delta ^3 (\mathbf P, \mathbf K)] = i\hbar \mathbf P \cdot \frac{\partial \delta^3 (\mathbf P, \mathbf K)}{\partial \mathbf P} , 
\end{equation}
between $\vert n \rangle, \vert m \rangle \in \{\vert n \rangle \}$. It is straightforward to show, with the help of the matrix representation (\ref{eq:ME_p_pp_delta_P}), that matrix elements of the commutator (\ref {eq:comm_G_delta3_P_K}) are defined on the space $D(\mathbb{R}^{3N})$. This is obtained in the representation in the basis $\vert \mathbf p_1, \dots, \mathbf p_N \rangle $. The vector subspace $D(\mathbb{R}^{3N})$ is a dense subspace of the Hilbert space  $L^2(\mathbb{R}^{3N})$. With the help of the Plancherel theorem for the Fourier transform on $L^2(\mathbb{R}^{3N})$ (\cite[Theorem IX.6]{4}), similar conclusion is inferred also for a dense subspace of the Hilbert space  $L^2(\mathbb{R}^{3N})$ obtained by Fourier transforms in the representation in basis $\vert \mathbf r_1, \dots, \mathbf r_N \rangle $.

From the relations (\ref{eq:ME_n_m_G_Proj_H_j2_Jz_Oa_Gauss_P}) and (\ref{eq:ME_n_m_Gauss_P_Proj_H_j2_Jz_Oa_G}) it follows that the limit $a \rightarrow 0$ of the matrix element of the commutator $[G,\phi _a(\mathbf P, \mathbf K)]$ in (\ref{eq:ME_n_m_G_comm_Gauss}) is the same if the projection operator $\mathcal {P}_{\mathcal {D}(H, J^2, J_{z}, \{\Omega ^\alpha \})}$ is inserted between the operators $G$ and $\phi _a (\mathbf P, \mathbf K )$. From the relation (\ref {eq:ME_G_comm_Gauss_P}), with the help of relation (\ref{eq:lim_Gauss_Sv_Sv}), one then obtains
\begin{equation}
\label{eq:ME_n_m_G_comm_dd_eq0_s} \langle n \vert G\mathcal{P}_{\mathcal {D}(H, J^{2}, J_{z}, \{\Omega ^\alpha \})}\delta ^3(\mathbf P,\mathbf K)  \vert m \rangle - \langle n \vert \delta ^3(\mathbf P, \mathbf K) \mathcal{P}_{\mathcal {D}(H, J^{2}, J_{z}, \{\Omega ^\alpha \} )} G \vert m \rangle = 0 .
\end{equation}
Matrix elements of the commutator (\ref {eq:comm_G_delta3_P_K}) between all $\vert n \rangle, \vert m \rangle \in \{\vert n \rangle \}$ are defined and are equal to zero. Since they are equivalent to relation (\ref {eq:ME_n_m_G_comm_dd_eq0_s}) each term in the relation (\ref {eq:ME_n_m_G_comm_dd_eq0_s}) is finite. From relations (\ref{eq:ME_dd_P_n_m}), (\ref{eq:F_T_fun_F}), (\ref{eq:F_n_m_k_Keq0}) and (\ref{eq:F_n_m_k_eq_F_n_m_l}) follows that the matrix elements $\langle n \vert \delta ^3 (\mathbf P, \mathbf K) \vert m \rangle $ between the eigenvectors $\{\vert n \rangle \}$ are unbounded if $\mathbf K = 0$. As stated in relation (\ref{eq:norm_delta_n}), operator $\delta ^3 (\mathbf P, \mathbf K=0)$ is not defined on the subspace $\mathcal {D}(H, \mathbf {J}^{2}, J_{z}, \{\Omega ^\alpha \})$ of the Hilbert space. 

The requirement that the matrix elements in relation (\ref {eq:ME_n_m_G_comm_dd_eq0_s}) are finite for all $\vert n \rangle $ and $\vert m \rangle$, and for  all $\mathbf K$ where $\mathbf K \in \mathbb{R}^3$, must be satisfied if the relation (\ref {eq:ME_n_m_G_comm_dd_eq0_s}) is valid. The only vector in $\mathcal {D}(H, \mathbf {J}^{2}, J_{z}, \{\Omega ^\alpha \})$ which is orthogonal to all vectors in $\mathcal {D}(H, \mathbf {J}^{2}, J_{z}, \{\Omega ^\alpha \})$ is the zero vector. Any linear operator acting upon a zero vector in some vector space maps it into a zero vector. Therefore, the only remaining possibility is that $\delta ^3 (\mathbf P, \mathbf K)$ acts upon the zero vector in each term in the relation (\ref{eq:ME_n_m_G_comm_dd_eq0_s}). The matrix elements in the relation (\ref{eq:ME_n_m_G_comm_dd_eq0_s}) are finite if and only if 
\begin{equation} 
\label{eq:Proj_H_J2_Jz_Oa_delta_p_n} \mathcal{P}_{\mathcal {D}(H, J^{2}, J_{z}, \{\Omega ^\alpha \} )} G \vert n \rangle = 0  ,
\end{equation}
for all $\vert n \rangle \in \{\vert n \rangle \}$. 

Projection operator $\mathcal{P}_{\mathcal {D}(H, J^{2}, J_{z}, \{\Omega ^\alpha \} )}$ is a bounded hermitian operator which maps every state vector onto its orthogonal projection in the subspace $\mathcal {D}(H, \mathbf {J}^{2}, J_{z}, \{\Omega ^\alpha \})$. From relation (\ref{eq:Proj_H_J2_Jz_Oa_delta_p_n}) follows immediately that the matrix elements of operator $G$ taken between all $\vert n \rangle, \vert m \rangle \in \{\vert n \rangle \}$ are equal to zero,   
\begin{eqnarray}
\label{eq:ME_n_m_G_eq0_all} \langle n \vert G \vert m \rangle = 0 , & \ \qquad \forall \vert n \rangle , \vert m \rangle \in \{\vert n\rangle \} \subset \mathcal {D}(H, \mathbf {J}^{2}, J_{z}, \{\Omega ^\alpha \}) . 
\end{eqnarray}
By applying the relation (\ref{eq:Proj_H_J2_Jz_Oa_delta_p_n}), or the relation (\ref{eq:ME_n_m_G_eq0_all}), the final result is obtained that the matrix elements of the commutator $[G, H] $, taken between all $\vert n \rangle, \vert m \rangle \in \{\vert n \rangle \}$ are equal to zero 
\begin{eqnarray}
\label{eq:ME_n_m_G_comm_H_eq0_D} \langle n \vert [G, H] \vert m \rangle = 0 , & \qquad \forall \vert n \rangle , \vert m \rangle \in \{\vert n\rangle \} \subset \mathcal {D}(H, \mathbf {J}^{2}, J_{z}, \{\Omega ^\alpha \}) . 
\end{eqnarray}

At the end of section the result is summarized in the following theorem:
\begin{tm}\label{theorem1}
Suppose that commutators of the Hamiltonian $H$ with the generators of translations and rotations, operators of total momentum $\mathbf {P}$ and total angular momentum $\mathbf {J}$, are equal to zero. Additional self-adjoint operators that commute with $H$ may also exist. Suppose that the additional self-adjoint operators which commute with $H$, $\mathbf J^2$, $J_z$ and among themselves, and hence are members of the set of commuting operators containing $H,\ \mathbf {J}^{2}$ and $\ J_{z}$, commute also with the generator of dilations $G$. These additional operators $\Omega ^\alpha $ form the set $\{\Omega ^\alpha \}$. Normalized simultaneous eigenvectors of the set of commuting operators $\{H$, $\mathbf {J}^{2}$, $J_{z}$, $\{\Omega ^\alpha\}\}$ which belong to the Hilbert space form an orthonormal basis in the subspace $\mathcal {D}(H, \mathbf {J}^{2}, J_{z}, \{\Omega ^\alpha \})$. Suppose that operator $G$ and all operators $P_iP_j\cdots P_n$, formed by the components $P_i$ of the vector operator $\mathbf P$, where $i,j,\dots, n = 1,2,3$, are defined on $\mathcal {D}(H, \mathbf {J}^{2}, J_{z}, \{\Omega ^\alpha \})$. Then the matrix elements of the commutator of the Hamiltonian $H$ with the generator of dilations $G$ are equal to zero on the subspace $\mathcal {D}(H, \mathbf {J}^{2}, J_{z}, \{\Omega ^\alpha \})$ of the Hilbert space.
\end{tm}

\section{Discussion}\label {sec:dilations}

Another approach to quantum mechanical virial theorem is based on scale transformation properties of the Hamiltonian. This approach uses the fact that operator $G$ is a generator of dilations, transformations which can be described as scalings of the position operators and inverse scalings of the momentum operators. Reference \cite{1} provides detailed explanations and derivations of the results that are possible by utilizing this approach to virial theorem. Some of the results taken from reference \cite{1} are now used here to clarify the main result of Sect. \ref {sec:theorem}.   

The proof of theorem in Sect. \ref {sec:theorem} is only formal, but avoids the difficulties exposed in \cite{5}. This is possible with the requirements of strong conditions in the theorem in Sect. \ref {sec:theorem}, which correspond to physically sensible properties. However, for any specific class of quantum mechanical systems that are invariant to translations and rotations, mathematically rigorous proof that all the conditions of this theorem are satisfied, or not, must be provided separately. This is a general characteristic of all forms of virial theorems known in the literature. If in this way, the applicability of this theorem is proved for a specific system, or for a class of systems, then it follows that for these systems the matrix elements of the commutator $[G,H]$ are equal to zero for all $\vert n \rangle, \vert m \rangle \in \{\vert n \rangle \}$. If in addition, the $N$-particle Hamiltonian has the form $H({\bf r}, {\bf p}) = T({\bf p}) + V({\bf r})$, then from relation (\ref{eq:ME_n_m_G_comm_H_eq0_D}), and also by using relations derived in reference \cite{1}, it follows that 
\begin{eqnarray} 
\label {eq:VT_dd} \langle n \vert \mathbf {p}\cdot {\partial T(\mathbf {p}) \over \partial \mathbf {p}} \vert n \rangle = \langle n \vert \mathbf {r}\cdot {\partial V(\mathbf {r}) \over \partial \mathbf {r}} \vert n \rangle , & \quad \forall \vert n \rangle \in \{\vert n\rangle \} \subset \mathcal {D}(H, \mathbf {J}^{2}, J_{z}, \{\Omega ^\alpha \}) . 
\end{eqnarray}
Here, $T(\mathbf {p})$ and $V(\mathbf {r})$ are the kinetic and potential energy operators, respectively; $\mathbf {p}$ denotes the set of $N$ momentum operators $(\mathbf {p}_{1}, \dots, \mathbf {p}_{1})$ and $\mathbf {r}$ denotes the set of $N$ position operators $(\mathbf {r}_{1}, \dots , \mathbf {r}_{N})$. Notation $\mathbf {p}\cdot {\partial  \over \partial \mathbf {p}}$ and $\mathbf {r}\cdot {\partial  \over \partial \mathbf {r}}$   stands for directional derivatives $\sum _{i=1}^{N} \mathbf {p}_{i} \cdot {\partial  \over \partial \mathbf {p}_{i}}$ and $\sum _{i=1}^{N} \mathbf {r}_{i} \cdot {\partial  \over \partial \mathbf {r}_{i}}$, respectively. It is implicitly assumed in relation (\ref {eq:VT_dd}) that the matrix elements of operators $\mathbf {p}\cdot {\partial T(\mathbf {p}) \over \partial \mathbf {p}}$ and  $\mathbf {r}\cdot {\partial V(\mathbf {r}) \over \partial \mathbf {r}}$ are defined on $\mathcal {D}(H, \mathbf {J}^{2}, J_{z}, \{\Omega ^\alpha \})$. 

Relation (\ref {eq:VT_dd}) is the quantum mechanical virial theorem that was already given in the form of an equivalent statement (\ref{eq:VT_sec2}) in Sect. \ref{sec:Vtfcaqs}. On the other hand, for a specific system for which the generalized virial theorem of Sect. \ref {sec:theorem} holds, a much stronger statement than just standard Eqs. (\ref{eq:VT_sec2}) and (\ref {eq:VT_dd}) is true. If an arbitrary state vector $\vert \psi \rangle$ in the subspace $\mathcal {D}(H, \mathbf {J}^{2}, J_{z}, \{\Omega ^\alpha \})$ generated by the basis $\{\vert n\rangle \}$ is expanded in this basis, i.e. if we write $\vert \psi \rangle = \sum _n c_n \vert n \rangle $, then the result 
\begin{eqnarray}
\label {eq:GVT_s_v} \langle \psi  \vert [G, H] \vert \psi  \rangle = 0 , 
\end{eqnarray}
is also obtained. Relation (\ref{eq:GVT_s_v}) is a generalization of the quantum mechanical virial theorem, on the subspace $\mathcal {D}(H, \mathbf {J}^{2}, J_{z}, \{\Omega ^\alpha \})$ of the Hilbert space. This is the main motivation behind this work and, if this result holds true for specific quantum mechanical systems, it should be discussed in the light of the current understanding of physical theories.

\section{Summary and perspectives}

This work is a generalization based on previously known facts on virial theorems in the nonrelativistic and relativistic quantum mechanics. It is demonstrated that this is attainable if certain requirements on symmetry properties of the Hamiltonian are made. The requirements are the conditions of the theorem in Sect. \ref {sec:theorem}. It is shown that if the conditions of translational and rotational symmetry of the Hamiltonian together with the additional conditions of the theorem in Sect. \ref {sec:theorem} are satisfied, the matrix elements of the commutator $[G, H]$ are equal to zero on the subspace $\mathcal {D}(H, \mathbf {J}^{2}, J_{z}, \{\Omega ^\alpha \})$ of the Hilbert space. Normalized simultaneous eigenvectors of the set of commuting operators $\{H$, $\mathbf {J}^{2}$, $J_{z}$, $\{\Omega ^\alpha\}\}$ which belong to the Hilbert space form an orthonormal basis in the subspace $\mathcal {D}(H, \mathbf {J}^{2}, J_{z}, \{\Omega ^\alpha \})$. 

The well known fact that operator $G$ is a generator of dilations, i.e. scaling transformations of the position operators and inverse scalings of the momentum operators, has been introduced in Sect. \ref {sec:dilations}. This approach is a basis for modern derivations of quantum mechanical virial theorems. In reference \cite{1} relation between directional derivatives of the kinetic and potential energy has been derived by extended use of the dilation approach and other approaches. Based on this, we consider here in Sect. \ref {sec:dilations} some implications for systems that satisfy the conditions of the theorem in Sect. \ref {sec:theorem}, and with the simple form of the $N$-particle Hamiltonian. For simplicity of presentation, more detailed forms of the Hamiltonian are not discussed here. Considering the complexity of $N$-body problems, it is author's expectation that further work on finding systems with the required properties will give more detailed answers.

Quantum mechanical virial theorem has proved important in large number of different areas. For example, the description of hadrons consisting of light quarks by two seemingly different approaches, in terms of the nonrelativistic Schr\"odinger formalism and by a semirelativistic Hamiltonian incorporating relativistic kinematics produces comparably good results. The relativistic generalization of the quantum mechanical virial theorem is derived and used to clarify the connection between the nonrelativistic and (semi)relativistic treatment of bound states in \cite{12}. It was also concluded in \cite{1} and \cite{7} that a massless particle described by the spinless Salpeter equation as well as a massless Dirac particle cannot be bound by a pure Coulomb potential. 

Other possibilities lead to reformulation of the virial theorem for quantum systems that are appropriate for a quantum field theoretic treatment. Based on the field theoretical canonical generator for the infinitesimal scale transformation of the second quantized Schr\"odinger field \cite {13}, a rigorous reformulation of the virial theorem for an interacting quantum many-body system with arbitrary spin is presented in \cite {14}. This formulation provides a general procedure applicable in the discussion on the equation of state in the framework of the nonperturbative canonical theory. A gauge invariant canonical generator for the scale transformation of the quantized Schr\"odinger field is proposed on the basis of the gauge invariance of the virial theorem in \cite{15}. In relativistic field theories scale transformations and virial theorem are more appropriately considered when the space coordinates and time are treated on the same ground. A relation between the trace anomaly of the energy-momentum tensor and the energy of a quantum bound state is obtained in \cite{16}. This anomaly is connected to the scale symmetry breakdown of quantum field theory.

\appendix* 
\section{Matrix representation of $\delta ^3(\mathbf P, \mathbf K)$} \label{app:1}

The limit $a \rightarrow 0$ of matrix element (\ref{eq:lim_Gauss_Sv_Sv}) of the function $\phi _a(\mathbf P, \mathbf K)$ introduced in Eq. (\ref{eq:Gauss_P_3D}), is considered here using two different bases. One of them is formed by the simultaneous eigenvectors $\vert \mathbf p_1, \dots, \mathbf p_N \rangle$ of the set of $N$ momentum operators $(\mathbf p_1, \dots, \mathbf p_N)$. The eigenvectors $\vert \mathbf p_1, \dots, \mathbf p_N \rangle $ satisfy $3N$ eigenvalue relations
\begin{equation}
p_{i l } \vert \mathbf p_1^\prime, \dots, \mathbf p_N^\prime \rangle = p_{i l}^\prime \vert \mathbf p_1^\prime, \dots, \mathbf p_N^\prime \rangle , 
\end{equation} 
where $p_{i l }$, with $l = 1,2,3$ , are components of the momentum operator $\mathbf p_i$ for the $i$-th particle. The basis vectors $\vert \mathbf p_1, \dots, \mathbf p_N \rangle$ are also eigenvectors of the total momentum operator $\mathbf P = \sum_{i=1}^{N} \mathbf p_i$, satisfying eigenvalue relations
\begin{equation}
P_l \vert \mathbf p_1^\prime, \dots, \mathbf p_N^\prime \rangle =  P_l ^\prime \vert \mathbf p_1^\prime, \dots, \mathbf p_N^\prime \rangle , 
\end{equation} 
where $P_l $, with $l = 1,2,3$, are components of the total momentum operator $\mathbf P$. 

The eigenvectors $\vert \mathbf p_1, \dots, \mathbf p_N \rangle$ are orthonormal in the sense defined by $\delta $-function normalization of $\vert \mathbf p_1, \dots, \mathbf p_N \rangle$, 
\begin{equation}
\langle \mathbf p_1 , \dots, \mathbf p_N \vert \mathbf {p}_1^\prime, \dots, \mathbf {p}_N^\prime \rangle = \prod _{i=1}^N \delta ^3(\mathbf p_i - \mathbf p_i^\prime) . 
\end{equation}
Matrix elements of the function $\phi _a(\mathbf P, \mathbf K)$ between the basis vectors $\vert \mathbf p_1, \dots, \mathbf p_N \rangle$ are obtained using the definition of $\phi _a(\mathbf P, \mathbf K)$ in (\ref{eq:Gauss_P_3D}),
\begin{equation}
\label{eq:ME_p_pp_Gauss} \langle \mathbf p_1 , \dots, \mathbf p_N \vert \phi _a(\mathbf P, \mathbf K) \vert \mathbf {p}_1^\prime, \dots, \mathbf {p}_N^\prime \rangle = \phi _a(\mathbf P^\prime , \mathbf K) \prod _{i=1}^N \delta ^3(\mathbf p_i - \mathbf p_i^\prime) .
\end{equation} 
Evaluating the matrix elements of the function $\phi _a(\mathbf P, \mathbf K)$ between the basis vectors $\vert \mathbf p_1, \dots, \mathbf p_N \rangle$, one obtains the function $\phi _a(\mathbf P, \mathbf K)$ of the total momentum eigenvalue $\mathbf P = \sum_{i=1}^{N} \mathbf p_i $. In relation (\ref{eq:ME_p_pp_Gauss}) this is indicated by a prime.

Relation (\ref{eq:ME_p_pp_Gauss}) is used in evaluating the limit $a \rightarrow 0$ of the matrix element of the function $\phi _a(\mathbf P, \mathbf K)$ taken between vectors $\vert \Psi \rangle $ and $\vert \Phi  \rangle $ in the Hilbert space, 
\begin{equation}
\label{eq:ps_psp_Gauss_lim} \lim _{a\rightarrow 0}\langle \Phi  \vert \phi _a(\mathbf P, \mathbf K) \vert \Psi \rangle = \lim _{a\rightarrow 0} \int \langle \Phi  \vert \mathbf p \rangle \phi _a(\mathbf P, \mathbf K) \langle \mathbf p \vert \Psi \rangle \prod _{i=1}^N d^3p_i .
\end{equation}
In relation (\ref{eq:ps_psp_Gauss_lim}) $\vert \mathbf p \rangle $ is a notation for $\vert \mathbf p_1, \dots, \mathbf p_N \rangle$. The function $\phi _a(\mathbf P, \mathbf K )$ on the right side of relation (\ref{eq:ps_psp_Gauss_lim}) corresponds to the delta distribution $\delta ^3(\mathbf P - \mathbf K) $. In accordance with the distributional definition of $\delta ^3(\mathbf x)$, introduced in Sect. \ref {sec:theorem}, relation (\ref{eq:ps_psp_Gauss_lim}) is written in the form
\begin{equation}
\label{eq:ps_psp_Gauss_delta_lim} \lim _{a\rightarrow 0}\langle \Phi  \vert \phi _a(\mathbf P, \mathbf K) \vert \Psi \rangle = \int \langle \Phi \vert \mathbf p \rangle \delta ^3(\mathbf P - \mathbf K) \langle \mathbf p \vert \Psi \rangle \prod _{i=1}^N d^3p_i .
\end{equation}
The following notation is introduced: 
\begin{equation}
\label{eq:ps_psp_delta_P} \lim _{a\rightarrow 0}\langle \Phi  \vert \phi _a(\mathbf P, \mathbf K) \vert \Psi \rangle = \langle \Phi  \vert \delta ^3(\mathbf P, \mathbf K) \vert \Psi \rangle  .
\end{equation} 
The limit $a \rightarrow 0$ of the matrix elements of the function $\phi _a(\mathbf P, \mathbf K)$ between vectors $\vert \Psi \rangle $ and $\vert \Phi \rangle $ in the Hilbert space in relation (\ref{eq:ps_psp_delta_P}) is evaluated in relations (\ref{eq:ps_psp_Gauss_lim}) and (\ref{eq:ps_psp_Gauss_delta_lim}). Furthermore, in the basis $\vert \mathbf p_1, \dots, \mathbf p_N \rangle$ the following notation is introduced: 
\begin{equation}
\label{eq:ME_p_pp_delta_P} \langle \mathbf p_1 , \dots, \mathbf p_N \vert \delta ^3(\mathbf P, \mathbf K ) \vert \mathbf {p}_1^\prime, \dots, \mathbf {p}_N^\prime \rangle = \delta ^3(\mathbf P^\prime - \mathbf K) \prod _{i=1}^N \delta ^3(\mathbf p_i - \mathbf p_i^\prime) .
\end{equation}
When it is possible, matrix elements (\ref{eq:ME_p_pp_delta_P}) are used equivalently with (\ref{eq:ME_p_pp_Gauss}) when evaluating the limit $a \rightarrow 0$ of matrix elements in relation (\ref{eq:ps_psp_delta_P}), as given by relations (\ref{eq:ps_psp_Gauss_lim}) and (\ref{eq:ps_psp_Gauss_delta_lim}).

Evaluation of this limit is possible in another basis formed by the simultaneous eigenvectors $\vert \mathbf r_1, \dots, \mathbf r_N \rangle$ of the set of $N$ position operators $(\mathbf r_1, \dots , \mathbf r_N)$. The eigenvectors $\vert \mathbf r_1, \dots, \mathbf r_N \rangle $ satisfy $3N$ eigenvalue relations
\begin{equation}
x_{i l } \vert \mathbf r_1^\prime, \dots, \mathbf r_N^\prime \rangle = x_{i l }^\prime \vert \mathbf r_1^\prime, \dots, \mathbf r_N^\prime \rangle , 
\end{equation} 
where $x_{i l }$, with $l = 1,2,3$ , are components of the position operator $\mathbf r_i$ for the $i$-th particle.
The eigenvectors $\vert \mathbf r_1, \dots, \mathbf r_N \rangle$ are orthonormal in the sense defined by $\delta $-function normalization of $\vert \mathbf r_1, \dots, \mathbf r_N \rangle$, 
\begin{equation}
\langle \mathbf r_1 , \dots, \mathbf r_N \vert \mathbf {r}_1^\prime, \dots, \mathbf {r}_N^\prime \rangle = \prod _{i=1}^N \delta ^3(\mathbf r_i - \mathbf r_i^\prime) . 
\end{equation}
The basis $\vert \mathbf r_1, \dots, \mathbf r_N \rangle$ and the basis  $\vert \mathbf p_1, \dots, \mathbf p_N \rangle$ are related by a unitary transformation 
\begin{equation}
\label{eq:UT_pb_rb} \vert \mathbf r_1 , \dots, \mathbf r_N \rangle = \int \vert \mathbf p_1 , \dots, \mathbf p_N \rangle \left (\mathbf {U}\right )_{\mathbf p, \mathbf r} \prod _{i=1}^N d^3p_i . 
\end{equation}
Elements of the unitary transformation matrix $\mathbf {U}$ are equal to:
\begin{equation} 
\left (\mathbf {U}\right )_{\mathbf p, \mathbf r} = \langle \mathbf p_1 , \dots, \mathbf p_N \vert \mathbf {r}_1, \dots, \mathbf {r}_N \rangle .
\end{equation}
Unitarity of the transformation matrix $\mathbf {U}$ ($\mathbf {U}^{\dag}\mathbf {U} = \mathbf {1}$ and $\mathbf {U}\mathbf {U}^{\dag} = \mathbf {1}$) follows from the completeness of orthonormal bases $\vert \mathbf p_1, \dots, \mathbf p_N \rangle$ and $\vert \mathbf r_1, \dots, \mathbf r_N \rangle$. The numbers of rows and columns of unitary transformation matrices $\mathbf {U}$ and $\mathbf {U} ^{\dag}$ are nondenumerably infinite. Each element $\left (\mathbf {U} \right )_{\mathbf p, \mathbf r} = \left (\mathbf {U} ^{\dag} \right )_{\mathbf r, \mathbf p}^{\ast }  $  of the matrices is a product of $N$ $\delta $-function normalized plane waves, 
\begin{equation}
\label{eq:pl_w_r_p} \left (\mathbf {U} \right )_{\mathbf p, \mathbf r} = \langle \mathbf p_1 , \dots, \mathbf p_N \vert \mathbf {r}_1, \dots, \mathbf {r}_N \rangle = \frac{1}{\left (2\pi \hbar\right)^{3N/2}  }\prod_{i=1}^N e^{- \frac{i}{\hbar } \mathbf p_i \cdot \mathbf r_i } .
\end{equation}

Finally, with the help of relation (\ref{eq:UT_pb_rb}), the matrix (\ref{eq:ME_p_pp_Gauss}) is transformed to the basis $\vert \mathbf r_1, \dots, \mathbf r_N \rangle$ by a unitary transformation  
\begin{equation}
\label{eq:ME_r_rp_p_pp_tr} \langle \mathbf r \vert \phi _a(\mathbf P, \mathbf K) \vert \mathbf {r}^\prime \rangle = \int \left (\mathbf {U} ^{\dag} \right )_{\mathbf r, \mathbf p}\langle \mathbf p \vert \phi _a(\mathbf P, \mathbf K) \vert \mathbf {p}^\prime \rangle \left (\mathbf {U}\right )_{\mathbf p^\prime , \mathbf r^\prime} \prod _{i=1}^N d^3p_i d^3p_i^\prime .
\end{equation} 
In relation (\ref{eq:ME_r_rp_p_pp_tr}) $\vert \mathbf r \rangle $ is a notation for $\vert \mathbf r_1, \dots, \mathbf r_N \rangle$ and $\vert \mathbf p \rangle $ is a notation for $\vert \mathbf p_1, \dots, \mathbf p_N \rangle$. With the help of relations (\ref{eq:ME_p_pp_Gauss}), (\ref{eq:pl_w_r_p}) and (\ref{eq:ME_r_rp_p_pp_tr}) one obtains
\begin{equation}
\label{eq:ME_r_rp_Gauss} \langle \mathbf r \vert \phi _a(\mathbf P, \mathbf K) \vert \mathbf {r}^\prime \rangle = \frac{1}{\left (2\pi \hbar\right)^{3N}  }\int \phi _a(\mathbf P, \mathbf K)\prod_{i=1}^N e^{\frac{i}{\hbar }\mathbf p_i \cdot \left (\mathbf r_i - \mathbf r_i^\prime \right ) } \prod _{j=1}^N d^3p_j .
\end{equation} 
The function $\phi _a(\mathbf P, \mathbf K )$ on the right side of relation (\ref{eq:ME_r_rp_Gauss}) corresponds to the delta distribution $\delta ^3(\mathbf P - \mathbf K) $. The limit $a \rightarrow 0$ of relation (\ref{eq:ME_r_rp_Gauss}) is therefore written in the form 
\begin{equation}
\label{eq:ME_r_rp_Gauss_delta_lim} \lim _{a \rightarrow 0} \langle \mathbf r \vert \phi _a(\mathbf P, \mathbf K) \vert \mathbf r^\prime \rangle = \frac{1}{\left (2\pi \hbar\right)^{3N}  }\int \delta ^3(\mathbf P - \mathbf K)\prod_{i=1}^N e^{\frac{i}{\hbar }\mathbf p_i \cdot \left (\mathbf r_i - \mathbf r_i^\prime \right ) } \prod _{j=1}^N d^3p_j . 
\end{equation}
With the help of the rules of $\delta $-calculus and the closure property of $\delta $-function normalized plane waves, one obtains from (\ref{eq:ME_r_rp_Gauss_delta_lim}) the following relation written in notation (\ref{eq:ps_psp_delta_P}), 
\begin{equation}
\label{eq:ME_r_rp_delta_P} \langle \mathbf r \vert \delta ^3(\mathbf P, \mathbf K) \vert \mathbf r^\prime \rangle = \frac{1}{\left (2\pi \hbar\right)^{3}  }e^{\frac{i}{\hbar }\mathbf K \cdot \left (\mathbf r_k - \mathbf r_k^\prime \right ) }\prod _{{i=1 \atop i\ne k}}^N \delta ^3(\mathbf r_i + \mathbf r_k^\prime - \mathbf r_k - \mathbf r_i^\prime ) .
\end{equation}
Index $k$ in relation (\ref{eq:ME_r_rp_delta_P}) can be any $k = 1, \dots , N$. Equivalently, relation (\ref{eq:ME_r_rp_delta_P}) is obtained directly by transforming the matrix (\ref{eq:ME_p_pp_delta_P}) to the basis  $\vert \mathbf r_1, \dots, \mathbf r_N \rangle$ by the same unitary transformation. Transforming the matrix (\ref{eq:ME_r_rp_delta_P}) to the basis $\vert \mathbf p_1, \dots, \mathbf p_N \rangle$ by inverse transformation, the matrix (\ref{eq:ME_p_pp_delta_P}) is again obtained.

\end{document}